# Band alignment study at the SrTaO$_2$N / H$_2$O interface varying lattice constants and surface termination from first-principles calculations

R. C. Bastidas Briceño,[1,2] V. I. Fernandez,[2,3] and R. E. Alonso[1,2,4]

[1] *Departamento de Ciencias Básicas, Facultad de Ingeniería Universidad Nacional de La Plata, La Plata, Argentina.*
[2] *Instituto de Física La Plata (IFLP), CONICET, Argentina.*
[3] *Departamento de Física, Facultad de Ciencias Exactas, Universidad Nacional de La Plata, Argentina..*
[4] *Instituto de Ingeniería y Agronomía, Universidad Nacional Arturo Jauretche, Argentina.*

**Abstract**

In the search for new renewable energy to replace fossil fuels, Hydrogen is one of the most promising candidates for clean energy production. But cheap Hydrogen separation and storage is still a big challenge. Photoelectrochemical devices look promising for the decomposition of the water molecule into $2H_2 + O_2$. Every day new materials and combinations are discovered or invented to improve the efficiency of the complex total process. A necessary condition for the photoelectrochemical process to work without a bias voltage is that the minimum of the semiconductor conduction band (CBM) must be more positive than the reduction potential $H^+$ to $H_2$, whereas the highest value of the semiconductor valence band (VBM) must be more negative than the oxidation potential of $H_2O$ to $O_2$. Thus, band alignment studies in interfaces of semiconductors with water become of vital importance.

In this work, first-principles calculations based on density–functional theory (DFT) in the all-electron and the pseudo-potential approaches have been performed for the analysis of the band alignment in SrTaO$_2$N/H$_2$O interfaces. Different surface terminations were analyzed, together with the dependence of the gap and band alignment with lattice constants for systems grown on mismatched substrates. Water structures were built from classical molecular dynamics and its electronic structure calculated using DFT. The calculations show that the SrTaO$_2$N (001) is suitable for photoelectrochemical applications on a wide range of lattice constant *a*, except for a compression/elongation of -2%, -1% and 3 %, while STN (110) results suitable for photoelectrochemical devices over a wider range of lattice constants from -1% to 3%.

Keywords: ab initio calculations, photoelectrochemical, hydrogen production

# 1. Introduction.

In the last decades many efforts have been made to discover new forms of clean, renewable and cheap energy to replace fossil fuels. Hydrogen is one of the most promising candidates as a next-generation energy carrier. Since the work of Honda and Fujishima in the early seventies [1], photocatalysis of water splitting using solar energy has become an active research area as a possible method of hydrogen production.

The water splitting reaction takes place when a semiconductor photocatalytic material surrounded by water molecules is irradiated with light with a photon energy equal or greater than its band gap. Then, electrons in the valence band (VB) are excited into the conduction band (CB), leaving holes in the VB. These photogenerated electrons and holes cause reduction and oxidation reactions on the water molecule, respectively. A necessary condition for the complete process of water splitting be fulfilled, is that the minimum of the semiconductor conduction band (CBM) must be more positive than the reduction potential $H^+$ to $H_2$, whereas the highest value of the semiconductor valence band (VBM) must be more negative than the oxidation potential of $H_2O$ to $O_2$. So, there is a minimum energy value of 1.23 eV needed for the reaction to be achieved. Although photocatalytic behavior was observed on many inorganic materials, conventional semiconductor photocatalytic materials present inadequate solar energy absorption (their gap is over 2.3 eV, while the desired band gap should be below 1.7 eV, the solar spectrum) and/or fast charge recombination events that result in a low efficiency for hydrogen production. [2].

Nowadays, the main goal in this research area is to find suitable photocatalytic materials with better efficiency (~ 10%) to be commercially viable. Perovskite-type oxynitrides $ABO_2N$ (A = Ca, Sr, Ba, La; B = Ti, Nb, Ta) are prospective materials for solar water splitting, mainly because they strongly absorb in the visible light spectrum, and some of they meet the necessary conditions for band edge alignments that lead to water splitting reactions. Unfortunately, several analyses of these materials report very poor catalytic activities. Among them, $SrTaO_2N$ has strong visible light absorption, but a poor photocatalytic activity. Recently, it was observed that when introducing Sr into a B site of the oxynitride [3] or synthesizing a solid solution between two oxynitrides $SrTaO_2N$ and $CaTaO_2N$ [4] the efficiency of the water splitting reaction in these materials was improved.

In a previous work, we analyzed how some properties of this material such as ferroelectricity can be modified when lattice constants slightly varied. Small variations on lattice constants can be achieved by growing a few layers of $SrTaO_2N$ onto a bulk of $SrTiO_3$, for instance. [5] Also, it should be noted that optical and photoelectrochemical properties will also change under tensile/compressive strain produced by thin films grown on different substrates. Thus, by doing this it should be in principle possible to tune the band alignment and gap, and therefore, change the natural properties of bulk $SrTaO_2N$ to enhance its photoelectrochemical performance. Also, the termination surface geometry is expected to play an important role in the latter. The aim of this work is to analyze the band alignment between perovskite-type oxynitride $SrTaO_2N$ (from now on STN) and water $H_2O$, considering different types of interfaces and varying STN lattice constants, and to observe if these variations might improve STN photocatalytic properties. The paper is organized as follows: In Section 2 the computational methods used for calculations are described. In Section 3 the system description and the configurations of the different structures and interfaces are detailed.

In Section 4 the band alignment method used is presented and the results are shown and discussed. Finally, Section 5 contains the summary and conclusions.

## 2. Computational method.

Calculations were performed in the framework of the Density Functional Theory (DFT) [6]. Firstly, for the gap study we used the all-electron full-potential linear augmented plane wave plus local orbital (FP-LAPW) method in the scalar relativistic version [7,9], by means of the WIEN2k implementation [10]. For the electronic exchange-correlation potential two different approximations were considered: the Perdew-Burke-Ernzerhof generalized gradient approximation (GGA) for solids (PBE-Sol) [11] and the modified Becke-Johnson exchange potential method (TB-mBJ).[12] As is known, the PBE-Sol approximation tends to underestimate the calculated GAP values, while TB-mBJ method produces a strong improvement, obtaining a better agreement with experimental results, and at the same time, comparable to calculations using hybrid functionals, but using considerably less computational resources. In this work, two versions as implemented in the WIEN2k code have been used and compared: TB-mBJ0 [12] and TB-mBJ1 [13]. The muffin-tin radii (RMT) were 2Å for Sr and Ta, and 1.65 Å for N and O. The parameter $RK_{Max}$ was set to 7, where $R_{MT}$ is the smallest muffin-tin radius and $R_{MT}K_{Max}$ is the largest wave number of the basis set. Integration in the reciprocal space was performed using the tetrahedron method using 300 k-points in the first Brillouin zone (which are reduced to 10 k-points in the irreducible wedge of the BZ).

Then, for the analysis of the structure properties and band alignment calculations we used the pseudopotential method as implemented by the QUANTUM ESPRESSO [14] code using GGA PBE-Sol. After several convergence tests and experimental data comparison, the following selections were made: for the electron-ion interaction the Rappe Rabe Kaxiras Joannopoulos ultrasoft pseudopotentials [15] were used, the wave functions were expanded by plane waves with a kinetic energy-cutoff of 80 Ry and for the charge density an energy-cutoff of 800 Ry. The irreducible Brillouin zone was sampled using the Monkhorst-Pack scheme with a 6x6x6 mesh [16].

## 3. System description

Bulk STN have been reported to exhibit perovskite cis-type configuration with I4/mcm space group at room temperature (RT) with lattice constants a= 5.70251(6) Å and c= 8.05420(16) Å [17] (See Fig. 1). The minimum energy structure predicted by DFT, obtained by varying cell constants and atomic positions is shown in Table 1. It can be observed to be a good agreement with the experimental data.

| Bulk SrTaO$_2$N | | | | | | | | |
|---|---|---|---|---|---|---|---|---|
| Variation | | -3% | -2% | -1% | optimized | 1% | 2% | 3% |
| Lattice Parameters Å | a | 5.478 | 5.535 | 5.591 | 5.648 | 5.704 | 5.761 | 5.817 |
| | c | 8.346 | 8.302 | 8.258 | 8.284 | 8.180 | 8.146 | 8.113 |

**Table 1: Lattice constant for the different elongated/compressed structures considered.**

In order to study STN and H$_2$O band alignment, a surface dividing these two specimens must be built. Then, on one side, a supercell was constructed, stacking *n* STN units, and then joining this supercell with a box containing liquid water (details of this box are explained later in this section). To establish an optimal number n of unit cells to be stacked, we performed several tests computing the difference between the mean values of the Hartree potential in the semiconductor (SC) side and the water side, to be less than 0.1 eV. This condition is fully accomplished for n=3.

For the SC side, the STN surface can be ended in different directions: in this work, we selected the principal two: the (001) and the (110) directions. Also, these two directions correspond to a polar-type (001) and nonpolar-type (110) of interfaces. Then, for each one of these structures, the in-plane lattice constants were varied to study the effect of in-plane tensile/compressive efforts. In this way, seven structures for the SC side for the (001) surface-termination were built: one corresponds to the equilibrium lattice constants *a= b,* and *c* obtained after optimization of the lattice constants and atomic positions (See Table 1). Then, a 2*a* x 2*a* x 3*c* supercell was built, as can be seen in Figure 2(a). The other six were obtained by varying +/- 3% the optimized lattice constant *a*, in 1% steps: -3%, -2%, -1%, 1%, 2% and 3%. Afterwards, the internal atomic positions and the lattice constant *c* were allowed to relax. At the same time, seven boxes with liquid water were modeled to be matched with each STN supercell. For each tension/compression value of *a*, the corresponding box presents m H$_2$O molecules in a box size of 2*a* x 2*a* x 13.17 Å (See Figure 2(b)). The number of molecules m was selected to approximately maintain the density of water as in normal temperature conditions: 1 g/cm$^3$. The same procedure was followed to build the seven structures for the (110) interface. Although in this case the lattice constant for the equilibrium structure were a*=$\sqrt{2}a$ , b* = c and c* = $\sqrt{2}b$, where *a* and *c* are the optimized bulk STN lattice constants. The supercells for the SC side were built as 1*a** x 1b *x 3*c**, and the corresponding liquid water boxes were built with box sizes of a* x b* x 13.17 Å.

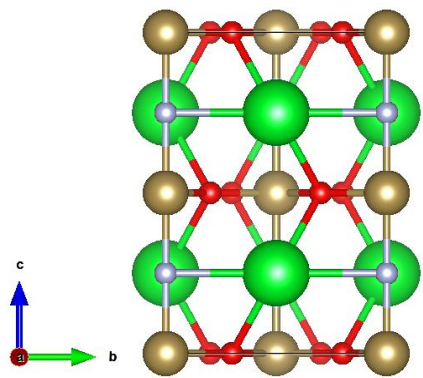

**Figure 1. Bulk SrTaO$_2$N**

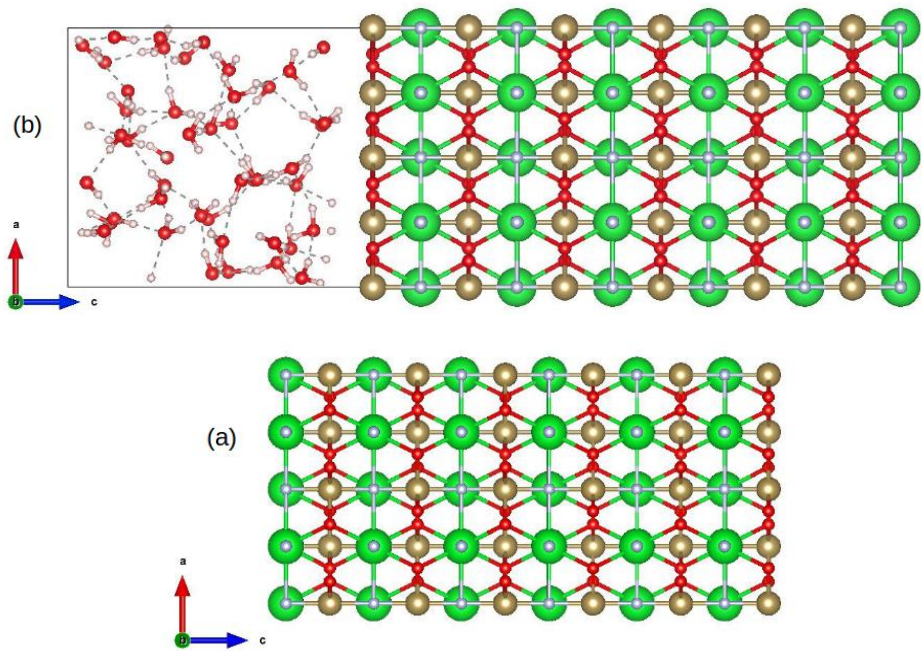

**Figure 2: Schematic atomic distribution in (001) 2x2x3-SrTaO$_2$N supercell (above) and (001) 2x2x3-SrTaO$_2$N /H$_2$O interface (below).**

Before implementing *ab-initio* calculations on the as-built STN/H$_2$O structures, the liquid water box configurations must be prepared. For that purpose, we started from an initial configuration of m H$_2$O molecules in a box, and then classical molecular dynamics (MD) was performed on it, using the DL_POLY program [18]. The interaction between water molecules was modelled with the TIP4P potential [19]. Then, NVT MD simulations during 100 ps were performed, and snapshots of the final atomic configurations were taken and used as input for the DFT calculations for each box. This whole procedure followed the previous work of Y. Wu [20]. As a test, for some of the so-obtained structures, Density of States (DOS) calculations were performed on the final atomic positions and compared with the DOS of corresponding systems whose atomic positions were obtained by quantum mechanical methods. In Figure 3, the DOS plot for a 56 molecules liquid-water system after 100 ps MD evolution and with DFT final relaxation is shown, which should be compared with figure 3 in Ref 20.

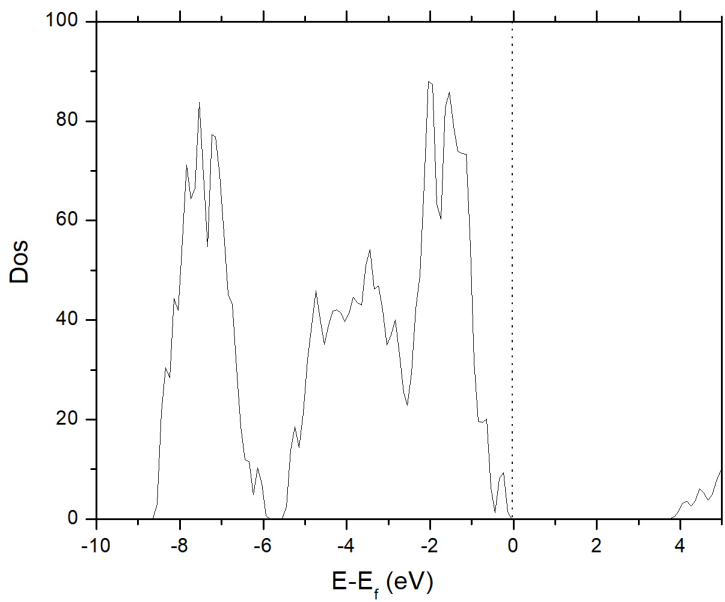

**Figure 3: Total DOS plot for a 56 molecules liquid-water system with atomic configurations obtained after 100 ps MD plus DFT final relaxation.**

## 4. Results and discussion.

STN presents an optical band gap of 2.1 eV [21] which, in principle, enables the water-splitting reaction without a bias voltage. Also, apart from an appropriate band gap, the photocatalyst material must satisfy that its CBM be more positive than the reduction potential of water and its VBM be more negative than the oxidation potential of water. If these conditions are fulfilled the water splitting reaction is energetically favorable without a bias voltage. Ab-initio calculations is a useful tool to study materials as prospective photocatalysts. One of the main problems is how to calculate the conduction and valence band positions of the semiconductor when it is in contact with water and compare them with the $H^+$ to $H_2$ and $H_2O$ to $O_2$ potentials in water. There are several methods in the literature proposing how to compute the band alignment. In this work, a three-step method proposed by Wu, Chang and Ceder, [20] was followed. In this method, the relevant energy levels (e.g., the CBM of the semiconductor, and the $H_2/H_2O$ level in water) at the interface are considered as the corresponding bulk energy levels subjected to the interfacial band bending effect. Wu et al. assume that band alignment is a consequence mainly of electrostatic effects, so the changes among the Hartree potential (HP) and the electronic energy levels stand nearly constant all over the space, and so does their difference. They demonstrate that the relative position between CB and $H_2/H_2O$ level at the interface can be computed by Eq. 1:

$$E_{CB(STN)}^{interface} - E_{H_2/H_2O}^{interface} = (E_{CB(STN)}^{bulk} - HP_{STN}^{bulk}) - (E_{H_2/H_2O}^{bulk} - HP_{water}^{bulk}) + (HP_{STN}^{interface} - HP_{water}^{interface})$$
(1)

The first term of the right part of Eq. 1, represents the CBM position relative to the average HP in STN bulk. The second term is the $H_2/H_2O$ level relative to the water's average HP in the water system and the third term is the difference between the average HP and STN and water in the interface system. Each of these three terms are determined by independent calculations:

1. $E_{CB(STN)}^{bulk} - HP_{STN}^{bulk}$: This term is calculated by running a variable cell relaxation on STN bulk, so the cell shape as well as the atomic coordinates are optimized.

2. $E_{H_2/H_2O}^{bulk} - HP_{water}^{bulk}$: This term represents the $H_2O/H_2$ acceptor level relative to the Hartree potential in the water system. To compute this term, it is necessary to calculate the lowest unoccupied molecular orbital (LUMO) level of water because this level is recognized as the acceptor level of water. These calculations can be performed with classical molecular dynamics (MD), replacing one molecule of $H_2O$ by a hydronium ion $H_3O^+$ in the water system. The whole calculation is detailed in Ref. [20] and is independent of the system under study because it is a water property, so the results obtained by Wu et al. will be used here.

3. $H_{STN}^{interface} - H_{water}^{interface}$: This last term is the outstanding one on the whole calculation and will be detailed in the following paragraphs.

For the VBM alignment, once the gap is known it can be computed by:

$$E_{VB(STN)}^{interface} - E_{H_2/H_2O}^{interface} = E_{CB(STN)}^{interface} + LUMO - GAP$$

(2)

| 001 Interfaces | | | | | | | | |
|---|---|---|---|---|---|---|---|---|
| Variation | | -3% | -2% | -1% | optimized | 1% | 2% | 3% |
| Lattice Parameters Å | 2a | 10.957 | 11.070 | 11.183 | 11.296 | 11.409 | 11.521 | 11.635 |
| | 3c | 36.123 | 36.003 | 35.880 | 35.770 | 35.952 | 35.572 | 35.481 |
| Number of molecules | $H_2O$ | 52 | 54 | 55 | 56 | 57 | 58 | 59 |
| length (Å) | a | 10.957 | 11.070 | 11.183 | 11.296 | 11.408 | 11.521 | 11.635 |
| height (Å) | c | 13.170 | 13.170 | 13.170 | 13.170 | 13.170 | 13.170 | 13.170 |
| Average HP without relax (eV) | STN | 2.45 | 2.48 | 2.46 | 2.39 | 2.37 | 2.37 | 2.26 |
| | $H_2O$ | 6.65 | 6.69 | 6.55 | 6.44 | 6.32 | 6.15 | 6.16 |
| Average HP with relax in z (eV) | STN | 1.92 | 1.98 | 1.99 | 1.83 | 1.84 | 1.65 | 1.58 |
| | $H_2O$ | 8.31 | 7.94 | 7.69 | 7.65 | 7.45 | 7.79 | 7.67 |
| 110 Interfaces | | | | | | | | |
| Lattice Paramters Å | a | 8.036 | 8.119 | 8.201 | 8.284 | 8.367 | 8.450 | 8.533 |
| | b | 7.748 | 7.828 | 7.907 | 7.987 | 8.067 | 8.147 | 8.227 |
| | c | 38.188 | 37.417 | 36.694 | 35.530 | 36.061 | 35.796 | 35.062 |
| Number of molecules | $H_2O$ | 26 | 26 | 27 | 27 | 28 | 28 | 29 |
| length (Å) | a | 8.036 | 8.119 | 8.201 | 8.284 | 8.367 | 8.450 | 8.533 |
| length (Å) | b | 7.748 | 7.828 | 7.907 | 7.987 | 8.067 | 8.147 | 8.227 |
| height (Å) | c | 13.170 | 13.170 | 13.170 | 13.170 | 13.170 | 13.170 | 13.170 |
| Average HP With relax in z (eV) | STN | 1.87 | 1.63 | 1.60 | 1.72 | 1.76 | 1.52 | 1.42 |
| | $H_2O$ | 7.41 | 7.49 | 7.44 | 7.44 | 7.42 | 7.38 | 7.39 |

**Table 2: (Upper) Cell Constants and average HP of the relaxed and unrelaxed STN/ $H_2O$ (001) interfaces. (Lower) Cell Constants and average HP of the relaxed STN/ $H_2O$ (110) interfaces.**

Figure 4 shows the diagram of the band alignment using the terminology of Eq. 1, and figure 5 shows, as an example, the HP obtained for the interface built with the cell constants and atomic positions obtained by the optimization of the STN in the (001) direction, along with their mean values calculated for each side of the interface. The calculated average HP values for STN and water at the interface for each case are presented in Table 2.

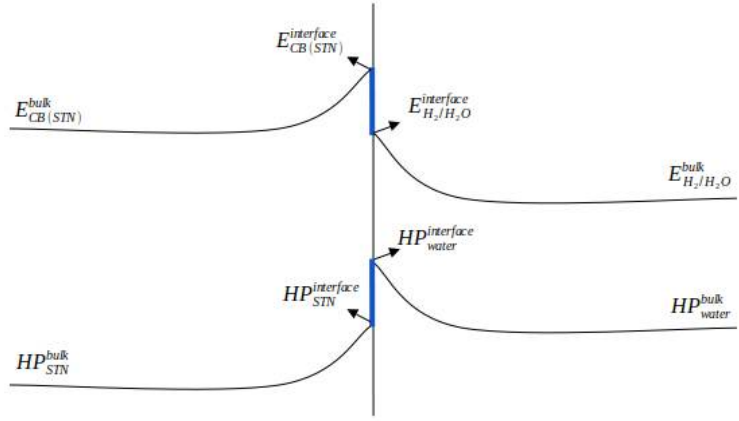

**Figure 4:** Band alignment diagram of $SrTaO_2N/H_2O$ interface: on the left side there are the STN bulk and surface bending CB band, and on the left side the corresponding to $H_2O$. The blue vertical lines represent the difference of the left side in Eq. 1. $E_{CB(STN)}^{bulk}$ = CBM in the bulk of the semiconductor. $E_{CB(STN)}^{interface}$ = CBM at $SrTaO_2N/H_2O$ interface, $E_{H_2/H_2O}^{bulk}$ = Bulk acceptor level ($H_2O$ / $H_2$ water level), $E_{H_2/H_2O}^{interface}$ = Acceptor level at $SrTaO_2N/H_2O$ interface. $HP_{STN}^{bulk}$ = HP on the bulk $SrTaO_2N$, $HP_{STN}^{interface}$ = HP on the semiconductor side at the $SrTaO_2N/H_2O$ interface. $HP_{water}^{bulk}$ = HP on the bulk $H_2O$, $HP_{water}^{interface}$ = HP on the solution side of the $SrTaO_2N/H_2O$ interface.

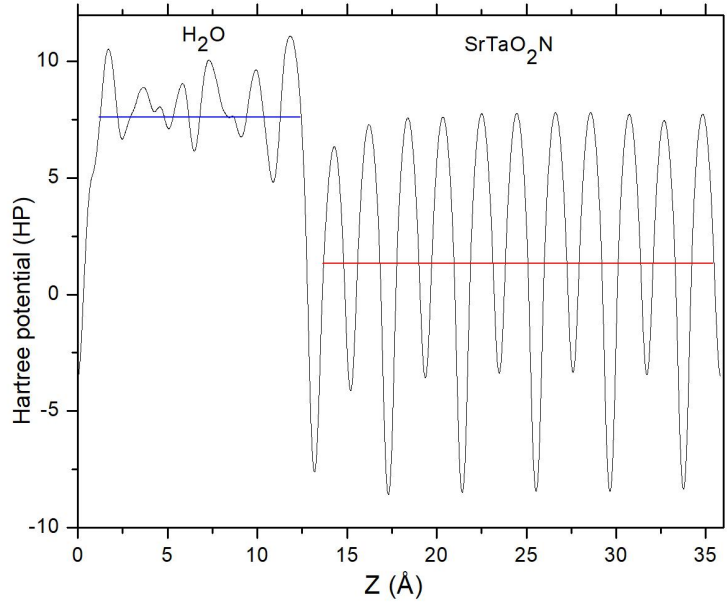

**Figure. 5.** Hartree potential at the interface optimized and relaxed the 001 direction. The blue line on the left shows the average Hartree potential for water, the red line on the right corresponds to the average Hartree potential for the semiconductor.

In this work, the behavior of the VB and CB was analyzed, in terms of band-alignment and gap size, for the different values of the cell constants parallel to the surface *a* and *b* previously mentioned, for two different terminations of the SC surface to observe their variation under tension/compression stress. In Table 3 the results obtained for the calculation of the gap in bulk STN as a function of the compression/stretching effort are shown. The experimentally reported value for unstressed STN is 2.1 eV [21]. For the unmodified cell, both TB-mBJ0 and TB-mBJ1 parameterizations are in good agreement with this experimental value, but TB-mBJ0 is closer. So, from now on, we will use results obtained with this parametrization.

| Structure | PBE [eV] | TB-mBJ0 [eV] | TB-mBJ1 [eV] |
|---|---|---|---|
| $a$-3% | 1.11 | 2.19 | 2.28 |
| $a$-2% | 1.04 | 2.20 | 2.29 |
| $a$-1% | 0.91 | 2.19 | 2.28 |
| $a$ | 0.90 | 2.17 | 2.25 |
| $a$+1% | 0.80 | 2.16 | 2.29 |
| $a$+2% | 0.61 | 2.02 | 2.16 |
| $a$+3% | 0.52 | 1.96 | 2.09 |

**Table 3: Calculated gap values on bulk STN with PBE and TB-mBJ as a function of lattice constant *a*.**

| 001 Interfaces | | | | | | | | |
|---|---|---|---|---|---|---|---|---|
| | | -3 % [eV] | -2 % [eV] | -1 % [eV] | optimized [eV] | 1% [eV] | 2% [eV] | 3% [eV] |
| without relax | CBM | 4.19 | 4.03 | 3.93 | 3.69 | 3.67 | 3.63 | 3.33 |
| | VBM | 2.00 | 1.83 | 1.74 | 1.54 | 1.50 | 1.61 | 1.37 |
| relaxed | CBM | 2.03 | 2.28 | 2.32 | 1.93 | 2.00 | 1.28 | 1.14 |
| | VBM | -0.16 | 0.08 | 0.13 | -0.23 | -0.15 | -0.75 | -0.82 |
| 110 Interfaces | | | | | | | | |
| relaxed | CBM | 2.85 | 2.38 | 2.18 | 2.03 | 1.81 | 1.56 | 1.26 |
| | VBM | 0.67 | 0.18 | -0.01 | -0.14 | -0.35 | -0.47 | -0.70 |

**Table 4. Band alignment as a function of lattice constant for (001) and (110) STN/ $H_2O$ interfaces.**

An important issue to consider is the importance of the surface effect, not only due to the shift of the bands corresponding to the atoms close to the surface, but also and in a decisive way, due to the displacement of these atoms because of the rupture of symmetry, as well as the eventual bonds with the water molecules on the other side of the surface. To analyze this, the process of calculating the alignment of the bands was repeated on the same structures, which previously underwent a new relaxation of the atoms belonging to the surface environment up to two atomic layers deep from the surface. Table 4, together with figures 6(a) and 6(b) show the results obtained of the alignment with and without the surface relaxation, for the (001) surface. As can be seen, the effect of the relaxation of the surface is decisive in the alignment, so for the case of the (110) surface only this case was considered. For this last, the obtained results are plotted in figure 7.

From the results shown in figures 6(b) and 7, it can be concluded that tensile/compressive effort, together with surface termination play a fundamental role on the gap and on the band alignment. STN (001) is suitable for photoelectrochemical applications except for a compression/elongation of its lattice constant *a* of -2%, -1% and 3 %. On the other hand, if epitaxial growth is performed on a substrate in such a way to build the surface on the 110 direction, STN results suitable for photoelectrochemical devices over a wider range of lattice constants from -1% to 3%. It can be seen from tables 3 and 4 and figure 7, that the elongation of the cell is accompanied by a decrease in the gap, together with a decrease in the CBM and VBM. In this way, depending on the growth method and the selection of the substrate, it would be possible to tune the alignment of the bands in order to optimize the photoelectrochemical process.

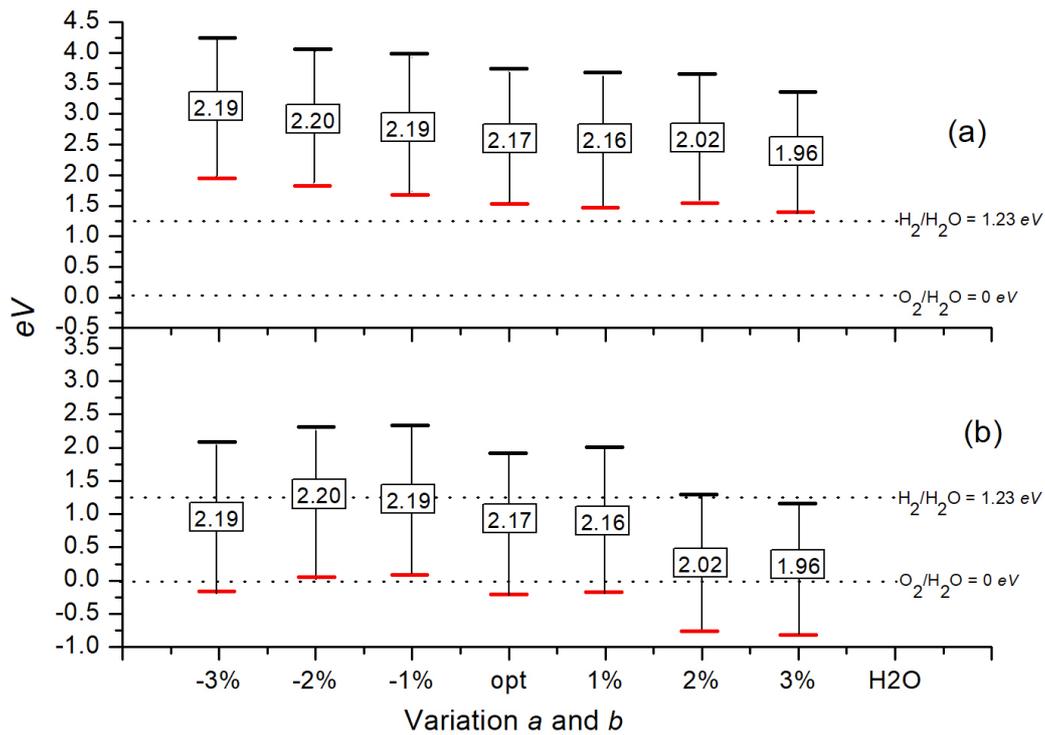

**Figure 6. Band alignment for the 001 structure. a) without relaxation of the atoms near the interface; b) with relaxation of the atoms near the interface. Red lines represent VBM and black lines CBM. The dotted lines indicate the $H_2O/H_2$ and $H_2O/O_2$ water levels. The rectangles show the obtained GAP values in eV using TB- mBJ0.**

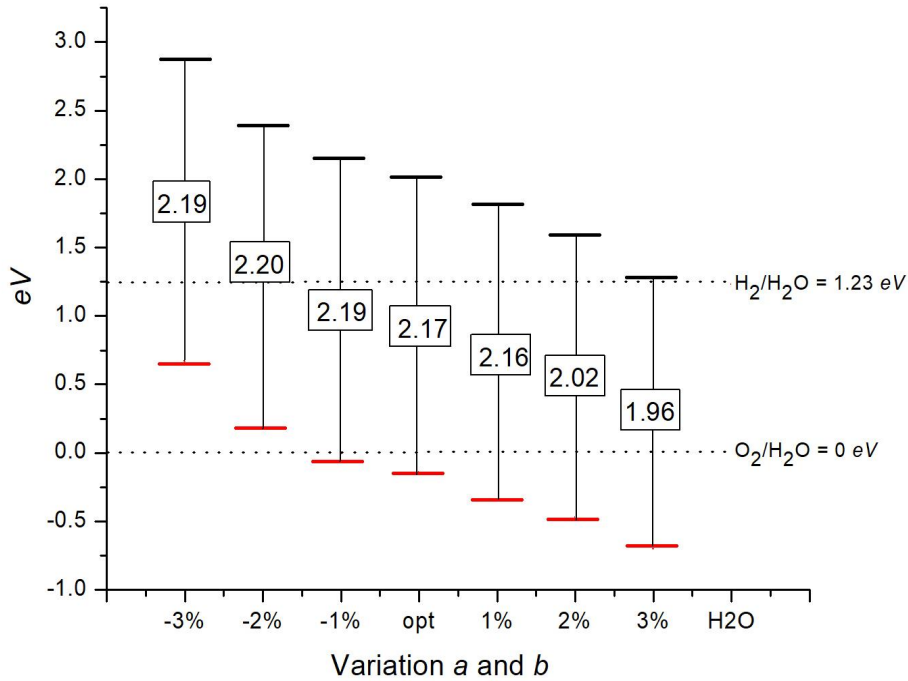

Figure 7: Band alignment for the 110 structures. Red lines represent VBM and CBM. black lines The dotted lines indicate the H₂O/H₂ and H₂O/O₂ water levels. The rectangles show the obtained GAP values in eV using TB- BJ0.

## 5. Summary and Conclusions

In this work, the behavior of the photoelectrochemical properties of STN subjected to tensile and compression stresses due to growth in thin layers onto mismatched lattice constant substrates was analyzed, through the behavior of the gap, the alignment of bands and the termination of the surface. Our results show the strong dependence of the alignment with the variation of the lattice constants. Also, the strong decrease of the gap in the case of stretching, and finally the different behavior depending on surface termination. According to the relative positions of the CBM and VBM of the SC and the water side, taking into account the band bending, STN (001) is suitable for photoelectrochemical applications on a wide range of lattice constant *a*, except for a compression/elongation of -2%, -1% and 3 %, while STN (110) results suitable for photoelectrochemical devices over a wider range of lattice constants from -1% to 3%. According to the analysis and tests carried out, the final relaxation of the atomic positions of the first and second shells on the STN side and that of all the atoms on the water side play a fundamental role in band alignment.

**Acknowledgments.**

This work was partially supported by Consejo Nacional de Investigaciones Científicas y Técnicas (CONICET) under Grants PIP11220170100987CO and PIO15520150100001CO, Facultad de Ingeniería, Universidad Nacional de La Plata under Grant I191 and Facultad de Ciencias Exactas, Universidad Nacional de La Plata under Grant X843. We also thank the computational centers CSCAA, Aarhus Universitet, Denmark, and Proyecto Acelerado de Cálculo of the SNCAD-MINCyT, Argentina. Also, to the Instituto de Física La Plata-CONICET, Argentina, for the use of its facilities. Finally, to Dra. M. A. Taylor and L. A. Errico for careful reading and inspiring discussions